# Nanoscopy of Excitons in Atomically Thin In-Plane Heterostructures with Nanointerfaces

Mahdi Ghafariasl[1,2], Tianyi Zhang[3], Sampath Gamage[1], Da Zhou[4], Muhammad Asjad[1], Sarabpreet Singh[1], Antonio Gómez-Rodríguez[6], Diego M. Solís[7], Venkataraman Swaminathan[8], Mauricio Terrones [3,4,5], Yohannes Abate [1*]

[1] Department of Physics and Astronomy, University of Georgia, Athens, GA, USA
[2] Joint Quantum Institute (JQI), University of Maryland, College Park, MD, USA
[3] Department of Electrical Engineering and Computer Science, Syracuse University, Syracuse, NY, USA
[4] Department of Physics, The Pennsylvania State University, University Park, PA, USA
[5] Department of Chemistry, The Pennsylvania State University, University Park, PA, USA
[6] Departamento de Tecnología de los Computadores y de las Comunicaciones, University of Extremadura, 10003 Cáceres, Spain
[7] Departamento de Teoría de la Señal y Comunicaciones, University of Vigo, 36301 Vigo, Spain
[8] Department of Materials Science and Nanoengineering, Rice University, Houston, TX, USA

**ABSTRACT**

Atomically sharp 2D in-plane heterostructures with nanointerfaces provide a powerful platform to tailor optical and electrical properties at the nanoscale, enabling novel device engineering and the exploration of new physical phenomena. However, direct experimental correlation between local dielectric response and excitonic properties across such interfaces has remained elusive. Here, we probed the nanoscale complex dielectric function and the corresponding localized photoluminescence (PL) modulations in heterostructure domains of lateral monolayer $Mo_xW_{1-x}S_2$ - $W_xMo_{1-x}S_2$, synthesized by a liquid-phase precursor-assisted approach. Near-field nanoimaging across the visible–near-infrared range enables real-space mapping of sharp amplitude and phase changes at the heterointerface, resolving the local complex dielectric function with nanometer-scale spatial resolution. Excitation-energy–dependent nano-spectroscopy reveals a reversal of dielectric contrast between Mo-rich and W-rich domains at their respective excitonic resonances, consistent with Lorentz-oscillator fits. Complementary hyperspectral nano-PL mapping resolves the evolution of excitonic emission across the lateral heterointerface, with neutral-exciton intensities varying continuously from W-rich to Mo-rich regions. Effective-medium theory modeling of the imaginary part of the effective dielectric function of the heterostructure as a function of photon energy and Mo filling fraction reproduces the observed excitonic trends, linking the PL evolution to a composition-dependent dielectric response. Low-temperature PL reveals exciton broadening in both Mo-rich and W-rich regions despite the presence of sharp interfaces, indicating defect-induced inhomogeneities compared to their pristine counterparts. Together, these results provide direct nanoscale correlation between dielectric and excitonic boundaries in laterally stitched monolayer heterostructures and establish a multimodal near-field spectroscopy framework for probing excitonic phenomena at the nanoscale.

**INTRODUCTION**

In-plane and out-of-plane heterostructures composed of atomically thin layers enable the design of excitonic systems with strong many-body effects[1-4]. Twisting or stitching different monolayers produces moiré patterns and emergent properties beyond those of individual layers[5,6]. In-plane heterostructures with lateral interfaces can be integrated into stacked architectures, with potential applications in field-effect transistors and planar p–n junctions[7-9]. In addition, interface features between dissimilar lateral domains—such as strain, defects, and alloying—strongly influence exciton behavior and oxidation, and provide a platform for exploring 1D quantum-confined many-body physics[10,11].

Unlike out-of-plane heterostructures that can be fabricated by manual van der Waals (vdW) stacking methods[12], the construction of in-plane heterostructures requires the formation of covalent bonds at the interface of two 2D material components, and thus needs to be achieved through controlled synthesis. Chemical vapor deposition (CVD) is a naturally suitable bottom-up approach for growing such in-plane heterostructures, because several growth parameters including the choices of precursors, temperature, pressure, and substrates, can be engineered to achieve the sequential growth of multiple 2D material components[13]. Especially, liquid-phase precursor-assisted CVD (i.e., the chalcogenation of liquid-phase metal precursors that are coated on the growth substrate) has emerged as a powerful technique, offering advantages such as cost-effectiveness, good scalability, and generalizability for synthesizing a wide range of 2D materials and heterostructures. It has enabled the construction of pristine and doped in-plane heterostructures with sharp 1D interfaces[14, 15].

Characterization of in-plane 2D heterostructures typically require a multimodal imaging and spectroscopy approach that combines structural microscopy with diffraction-limited optical microscopy and spectroscopy. Atomic-resolution techniques, such as scanning transmission electron microscopy (STEM) and aberration-corrected high-resolution transmission electron microscopy (AC-HRTEM), are widely used to map composition, dopant, and alloy distributions, with AC-HRTEM offering enhanced imaging capability[16]. Chemical composition and excitonic properties can be characterized by far-field optical methods such as Raman and photoluminescence (PL) spectroscopy that can reveal how in-plane concentration differences, defects, doping, and alloying impact optical and electronic properties [17]. High spatial resolution methods are essential for studying nanointerfaces, which conventional far-field optical techniques cannot provide due to the diffraction limit, preventing resolution of features smaller than half the illumination wavelength. Tip-based inelastic scattering techniques, such as tip-enhanced photoluminescence (TEPL) and tip-enhanced Raman spectroscopy (TERS), overcome this limitation by nano-confining and enhancing the electric field at the tip apex, enabling optical studies with sub-diffraction spatial resolution [18-22]. A powerful tip-based elastic scattering technique, scattering-type scanning near-field optical microscopy (s-SNOM), provides compositional and spectroscopic imaging at nanoscale resolution, independent of the illumination wavelength, across the electromagnetic spectrum[23].

In this work, we synthesize an in-plane heterostructure consisting of a Mo-rich center and W-rich edge using the aforementioned liquid-phase precursor-assisted CVD method, forming a sharp 2D alloyed interface. Aberration-corrected high-resolution scanning transmission electron microscopy (AC-HRSTEM) reveals that the center and edge are primarily $MoS_2$ and $WS_2$, respectively, each doped with trace amounts of the other metal, and the interface consists of an alloyed $Mo_xW_{1-x}S_2$ region, which we refer to collectively as $Mo_xW_{1-x}S_2$ - $W_xMo_{1-x}S_2$ s-SNOM was used to map the heterostructure and its interface, revealing well-defined compositional boundaries: near-field images show abrupt contrast changes across the junction, from which the interface sharpness was determined to be less than 70 nm. Importantly, the normalized near-field amplitude and phase signals measured at the Mo-rich center and W-rich edge are in good agreement with the corresponding local complex dielectric function $\varepsilon(\omega)$. AFM-based PL spectroscopy at room temperature reveals spatial variations in exciton peak position and intensity across $MoS_2$-like and $WS_2$-like regions, indicating sharp optical contrast at the in-plane interface. Low-temperature PL measurements reveal significant exciton linewidth broadening in both the Mo-rich and W-rich regions, despite the presence of nano-sharp interfaces. This broadening, absent in pristine monolayers, points to underlying disorder such as compositional variations, strain fields, or defect states introduced during synthesis. These results highlight the potential of lateral 2D heterostructures for nanoscale optical control, offering a platform for next-generation quantum confined device engineering.

**RESULTS AND DISCUSSION**

A liquid-phase precursor-assisted synthesis approach, which we have developed previously [24, 25], was used to produce the monolayer in-plane heterostructures (see Materials and Methods section). This method involves the spin-coating of uniformly mixed water-soluble Mo- and W-precursors and the growth promoter (sodium cholate, $C_{24}H_{39}NaO_5$) onto a $SiO_2$/Si wafer, followed by a sulfurization process inside a tube furnace (Fig. 1(a)). An optical image of typical as-grown in-plane heterostructures is displayed in Fig. 1(b), showing triangular morphologies with distinct optical contrast between center and edge regions. According to our previous Z-contrast aberration-corrected high-angle annular dark field scanning transmission electron microscopy (AC-HAADF-STEM) observation on the sample synthesized under identical experimental conditions, the resultant in-plane heterostructures possess a Mo-rich center and a W-rich edge region, with an alloyed 2D interface [24] (a schematic of the configuration of heterostructures is shown in Fig. 2(a)). To verify the spatial resolved compositional difference, we performed AC-HRSTEM characterization. Figures 1(c)-(e) show high-angle annular dark field (HAADF)-STEM images obtained from center, interface, and edge regions of a typical flake. In the HAADF imaging mode, the image intensity of atoms and their atomic Z-numbers have a positive correlation, and thus Mo atoms display darker contrast due to their relative lower Z-number which can be differentiated from W atoms[26, 27]. Thus, it can be deduced from Fig. 1(c)-(e) that the center and edge regions are predominantly $MoS_2$ and $WS_2$ with a small amount of W and Mo dopants, respectively, and the interface region is alloyed $Mo_xW_{1-x}S_2$, in agreement with our previous observations. Due to the presence of dopants in each component of the in-plane heterostructures, we use the name $Mo_xW_{1-x}S_2$ - $W_xMo_{1-x}S_2$ for the heterostructure sample studied in this work[24].

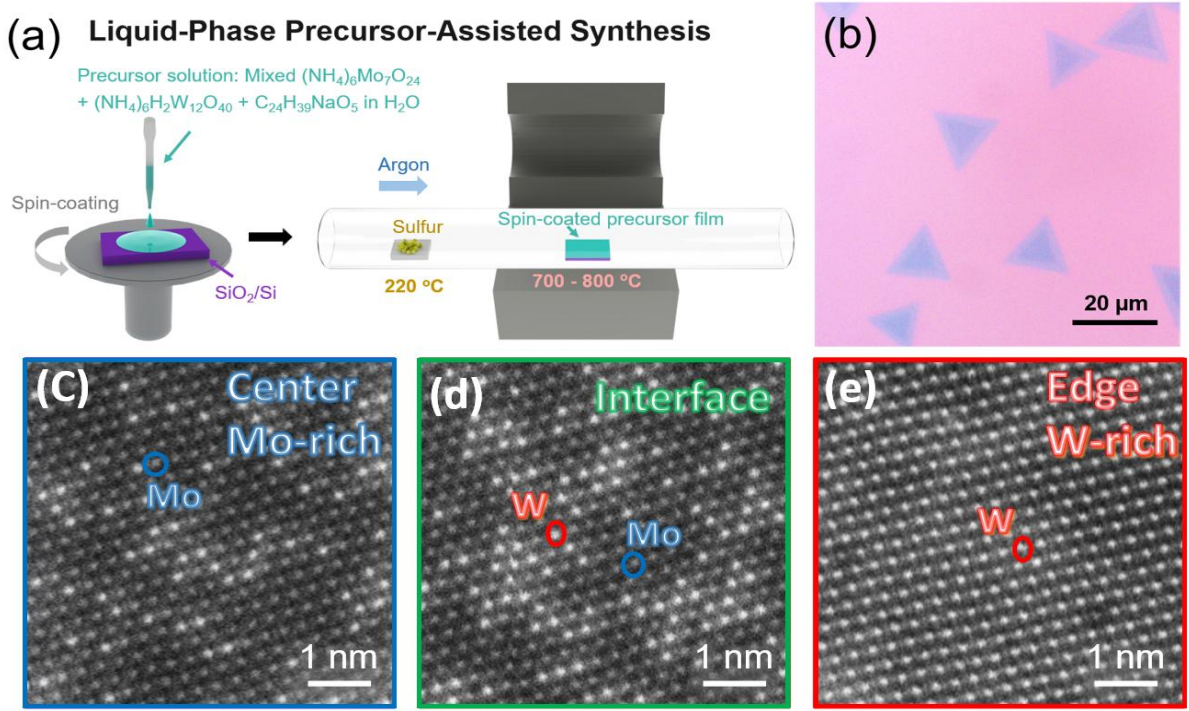

**Figure 1. Synthesis and AC-HAADF-STEM structural characterization of monolayer $Mo_xW_{1-x}S_2$ - $W_xMo_{1-x}S_2$ in-plane heterostructures.** (a) A schematic of the liquid-phase precursor-assisted synthesis and (b) an optical image of $Mo_xW_{1-x}S_2$ - $W_xMo_{1-x}S_2$ in-plane heterostructures on $SiO_2/Si$ substrate. (c-e) Representative AC-HAADF-STEM images acquired from the center, interface, and edge areas of the heterostructures.

We performed far-field Raman and PL measurements to study the structural and optical properties of as-synthesized in-plane heterostructures that are schematically shown in Fig. 2(a). Figure 2(b) illustrates the Raman spectra acquired from different regions of the $Mo_xW_{1-x}S_2$ - $W_xMo_{1-x}S_2$ in-plane heterostructures. The material exhibits similar Raman signatures with previous results, with $MoS_2$-like vibrational modes dominating in center regions and $WS_2$-like vibrational modes in edge regions, while the interface displays a combination of both sets of modes [24, 28]. The PL emissions peaks from the center (~1.82 eV) and edge regions (~1.96 eV) are close to the optical band gap of pristine $MoS_2$ and $WS_2$, respectively[29, 30]. At the interface of the in-plane heterostructures, two separate peaks, corresponding to $MoS_2$-like PL and $WS_2$-like PL, were observed (Fig. 2(c)). The PL result corroborates the existence of a sharp composition transition within the material, indicating a Mo-rich center and W-rich edge regions which allows two distinct PL peaks to be observed upon laser excitation. This PL feature marks a distinct difference between our in-plane heterostructures and previously reported alloyed $Mo_xW_{1-x}S_2$ with a composition gradient[17], as in the case of the alloy, a single peak is consistently displayed at different locations due to the absence of a sharp interface.

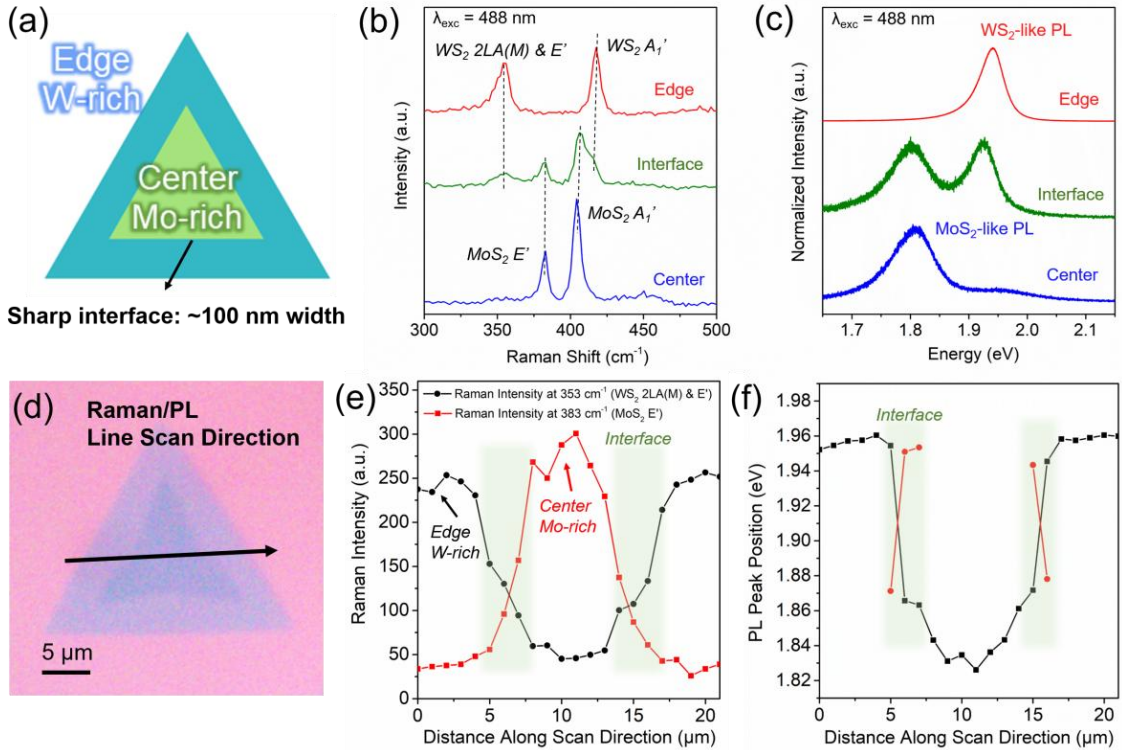

**Figure 2. Optical characterization of monolayer $Mo_xW_{1-x}S_2$ - $W_xMo_{1-x}S_2$ in-plane heterostructures.** (a) Schematic of the structure of monolayer $Mo_xW_{1-x}S_2$ - $W_xMo_{1-x}S_2$ in-plane heterostructures. (b) Raman and (c) PL spectra obtained from center, interface, and edge regions of the in-plane heterostructures. (d) Optical microscopy image of the monolayer $Mo_xW_{1-x}S_2$ - $W_xMo_{1-x}S_2$ flake used for Raman and PL line scans (the scan direction is marked with the black arrow). (e) Raman intensity profiles of $WS_2$ E' and 2LA(M) modes (black) and $MoS_2$ E' mode (red line) as a function of distance along the scan direction. Both Raman modes display dramatic changes at the interface (marked with the green box). (f) The evolution of PL peak positions of the in-plane heterostructures as a function of distance along the scan direction. The dramatic shift in the peak position along the scan direction indicates the presence of hetero-interfaces (marked with the green box).

To further elucidate the structural and optical property as a function of position within the flake, Raman and PL line scans were subsequently performed along the line-scan direction marked by the arrow in Fig. 2(d). Raman intensity line scans of WS$_2$ E' and 2LA(M) modes (black line corresponding to the 353 cm$^{-1}$ peak in panel (b)) as well as MoS$_2$ E' mode (red line corresponding to the 380 cm$^{-1}$ peak in panel (b)) of the Mo$_x$W$_{1-x}$S$_2$ - W$_x$Mo$_{1-x}$S$_2$ in-plane heterostructures are shown in Figure 2(e). As expected, the edge (center) regions are dominated by Raman modes of WS$_2$ (MoS$_2$). The regions with coexisting MoS$_2$- and WS$_2$-Raman modes were highlighted in green, which represent interface regions. We then performed PL line scans along the identical scan direction on the same flake and plotted the PL peak positions as a function of location (Fig. 2(f)), which showed WS$_2$-like PL emission energy (~1.94-1.96 eV) at edge regions, and MoS$_2$-like PL emission energy (~1.82-1.86 eV) at center regions. Similar to the Raman line scan results, there are interface regions with both MoS$_2$- and WS$_2$-like PL peaks (highlighted in green). It should be noted that the interface width (~ 4 µm) observed in Raman and PL line scans is much larger than TEM-measured result (~20-100 nm), due to the larger laser spot size compared to the interface width that limits the resolution of far-field Raman and PL line scans, and thus the interface cannot be clearly resolved. Nevertheless, the clear WS$_2$-to-MoS$_2$ transition from edge to center regions, observed in both Raman and PL line scans, confirms the existence of interfaces within the sample.

We employed near-field nanoimaging in the visible and near infrared frequency range to map the dielectric function of the heterostructure which enabled us to distinguish the different regions including the sharp interface boundary of the in-plane heterostructure. We coupled an ultra-stable and tunable continuous-wave Ti-sapphire laser source to a s-SNOM system (Attocube GmbH) and acquired demodulated 4$^{th}$ harmonic near-field amplitude and phase images of the monolayer heterostructure. The illumination laser is tuned to energies 1.72, 1.75, and 1.97 eV, to visualize the W-rich and Mo-rich domains distinctly. Figure 3(a) depicts the surface topography of the monolayer Mo$_x$W$_{1-x}$S$_2$-W$_x$Mo$_{1-x}$S$_2$ sample, which identifies the inner and outer regions of the sample. Panels (b) and (c) in Fig. 3 present the 4$^{th}$ harmonic amplitude (A$_4$) and phase (φ$_4$) images, respectively at three different excitation energies. The amplitude and phase images reveal sharp image contrast transitions across the interface, highlighting the heterostructure's well-defined compositional boundaries. The monochromatic near-field images reveal the local complex permittivity (ε) of the sample. The normalized amplitude signals correspond to the real part (Re(ε)) of the permittivity, while the normalized phase signals represent the imaginary part (Im(ε))[23]. In the amplitude images at 1.72 eV, the Mo-rich central region appears brighter (yellow) than the W-rich edges, whereas at 1.97 eV, the contrast reverses, with the W-rich edges becoming brighter. Both the W-rich and SiO$_2$ substrate appear as dark (red) in the amplitude images at 1.72 and 1.75 eV since the near-field amplitude is smaller than the brighter MoS$_2$ sample, because the W-rich and SiO$_2$ have a smaller permittivity than MoS$_2$ at the imaging frequency. Likewise, the phase images reveal contrast associated with the imaginary part of the permittivity. At 1.72 eV and 1.75 eV, a clear distinction is observed between the center and edge regions, whereas at 1.97 eV, this distinction vanishes. This energy-dependent contrast reflects material-specific excitonic transitions, where WS$_2$ exhibits a strong absorption peak at 1.97 eV, while MoS$_2$ dominates at lower energies (1.72 and 1.75 eV), consistent with Lorentzian dielectric models for TMDCs[31]. The line profiles plotted in Figures 3(d)(i & ii) further illustrate the spatial variation of the near-field amplitude and phase signals across the Mo$_x$W$_{1-x}$S$_2$-W$_x$Mo$_{1-x}$S$_2$ in-plane heterostructure. The blue and black connecting curves exhibit a similar trend across the sample, while the red curve shows a smaller amplitude and phase signal in the Mo-rich region compared to the W-rich region. e, To complement the experimental near-field results, we parameterized the complex dielectric response of monolayer MoS$_2$ and WS$_2$ using a weak Lorentz-oscillator model consistent with Kramers–Kronig relations[9,23,31]. The real ($\varepsilon_1$) and imaginary ($\varepsilon_2$) parts of the dielectric function were calculated over the 0.5–2.35 eV range using a set of Lorentz terms that reproduce the characteristic A, B excitonic resonances reported by Li *et al.*[9,31] As shown in Fig. 3(e), the complex permitivities capture the dispersive and absorptive behavior of both materials: MoS$_2$ exhibits strong $\varepsilon_2$ peaks near 1.9 eV and 2.1 eV corresponding to its A and B excitons, while WS$_2$ shows higher-energy resonances at 2.06 eV. The corresponding $\varepsilon_1$ curves display the expected dispersive sign reversals around these excitonic energies. These simulations reproduce the qualitative dielectric trends observed in the experimental near-field amplitude and phase contrasts (Fig. 3b–d), providing a self-consistent link between the measured optical response and the intrinsic excitonic dielectric properties of each domain. Using the near-field images shown in Fig. 3(b,c), we extracted quantitative fourth-harmonic near-field amplitude and phase signals from identical spatial regions of interest within the Mo-rich and W-rich domains. The signals were normalized to the SiO$_2$ substrate response and evaluated as a function of excitation energy. The resulting experimental near-field amplitude and phase values are shown as discrete markers in Fig. 3(e), overlaid on the simulated dielectric functions. The solid curves in Fig. 3(e) represent the real ($\varepsilon_1$) and imaginary ($\varepsilon_2$) parts of the dielectric function of monolayer $MoS_2$ and $WS_2$ calculated using a weak Lorentz-oscillator model, while the black and red circular markers correspond to the experimentally extracted fourth-harmonic near-field amplitude and phase signals from the selected regions. The experimental data shows excellent agreement with the simulated dielectric response.

At the Mo-rich excitonic resonance (≈1.72 eV), the real part of the dielectric function $\varepsilon_1$ is larger in the Mo-rich region than in the W-rich region, resulting in an enhanced near-field amplitude in the Mo-rich domain. Conversely, at the W-rich resonance (≈1.97 eV), the trend reverses, with the W-rich region exhibiting a larger $\varepsilon_1$ and a correspondingly higher near-field amplitude, as shown in Fig. 3(e)(i). A similar inversion is observed in the phase response, which tracks the imaginary component $\varepsilon_2$ of the dielectric function, as shown in Fig. 3(e)(ii). This strong correspondence between the experimentally measured near-field signals and the simulated dielectric functions confirms that the observed near-field contrast directly reflects the local excitonic dielectric response of each domain in the lateral heterostructure.

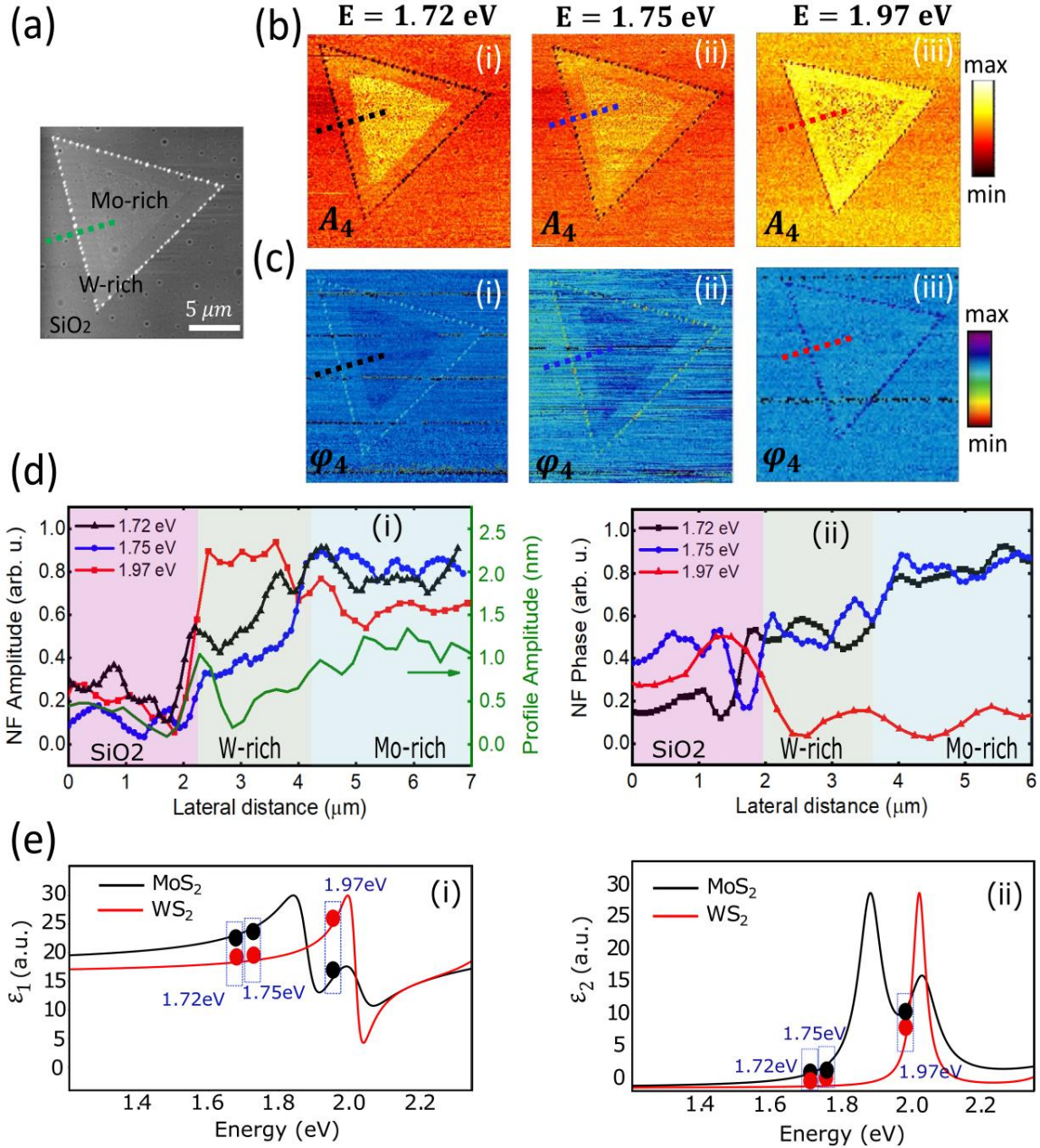

**Figure 3. Near-field optical response of the monolayer Mo$_x$W$_{1-x}$S$_2$ - W$_x$Mo$_{1-x}$S$_2$ heterostructure at different excitation energies.** (a) AFM topography of the monolayer heterostructure, showing Mo-rich and W-rich regions on a SiO$_2$ substrate. (b,c) 4$^{th}$ harmonic near-field amplitude (A$_4$) and phase (φ$_4$) images at illumination laser energies 1.72 eV, 1.75 eV, and 1.97 eV revealing pronounced energy-dependent contrast across the lateral heterointerface. Dashed lines indicate the positions used for extracting line profiles. (d) (i) Near-field amplitude line profiles extracted along the dashed lines in panels (b,c) at different excitation energies, overlaid with the corresponding AFM height profile. (ii) Near-field phase line profiles across the same regions, highlighting the contrast inversion between Mo-rich and W-rich domains as a function of energy. (e) Simulated real (ε$_1$) and imaginary (ε$_2$) parts of the dielectric function for monolayer MoS$_2$ (black curves) and WS$_2$ (red curves) using the weak Lorentz-oscillator model in the photon-energy range of 0.5–2.35 eV. The solid curves represent the modeled dielectric response, while the discrete black and red circular markers correspond to experimentally extracted quantitative fourth-harmonic near-field amplitude and phase values, normalized to the SiO$_2$ substrate, at the excitation energies used in panels (b–d).

Figure 4 shows high-resolution s-SNOM imaging of the interface in the Mo$_x$W$_{1-x}$S$_2$-W$_x$Mo$_{1-x}$S$_2$ in-plane heterostructure. Figure 4 (a) shows the topography (i) and the 4$^{th}$ harmonic amplitude (A$_4$) and phase ($\varphi_4$) maps (ii and iii, respectively) at 1.72 eV, revealing distinct optical contrasts that highlight the sharpness of the boundary of W-rich and Mo-rich regions. Panel (b) provides the line profiles of the 4$^{th}$ harmonic amplitude ($A_4/A_{4,SiO2}$) and phase ($\varphi_4 - \varphi_{4,SiO2}$) across the interface region. The amplitude ratio (red dashed line) and phase difference (blue dashed line) clearly show distinct optical responses for the W-rich and Mo-rich regions, showing permittivity variations between the two domains as discussed above. Panel (c) shows the amplitude (A$_4$) line profile taken along the dashed line at the interface region. The black dashed line represents experimental data, while the solid red line is the fitted curve, providing a smooth transition between the W-rich and Mo-rich regions. The spatial resolution was calculated by taking the full width at half maximum (FWHM) of the line spread function (LSF) (Fig. 4(d)), which is obtained by the derivative of the amplitude line profile in Fig. 4(c) [32, 33]. By fitting this derivative of the amplitude line profile with a symmetric Lorentzian curve (red line), the FWHM is determined to be 67 nm. This provides direct experimental evidence of the sharpness of the Mo-W interface which is in the range of the size of the probe tip.

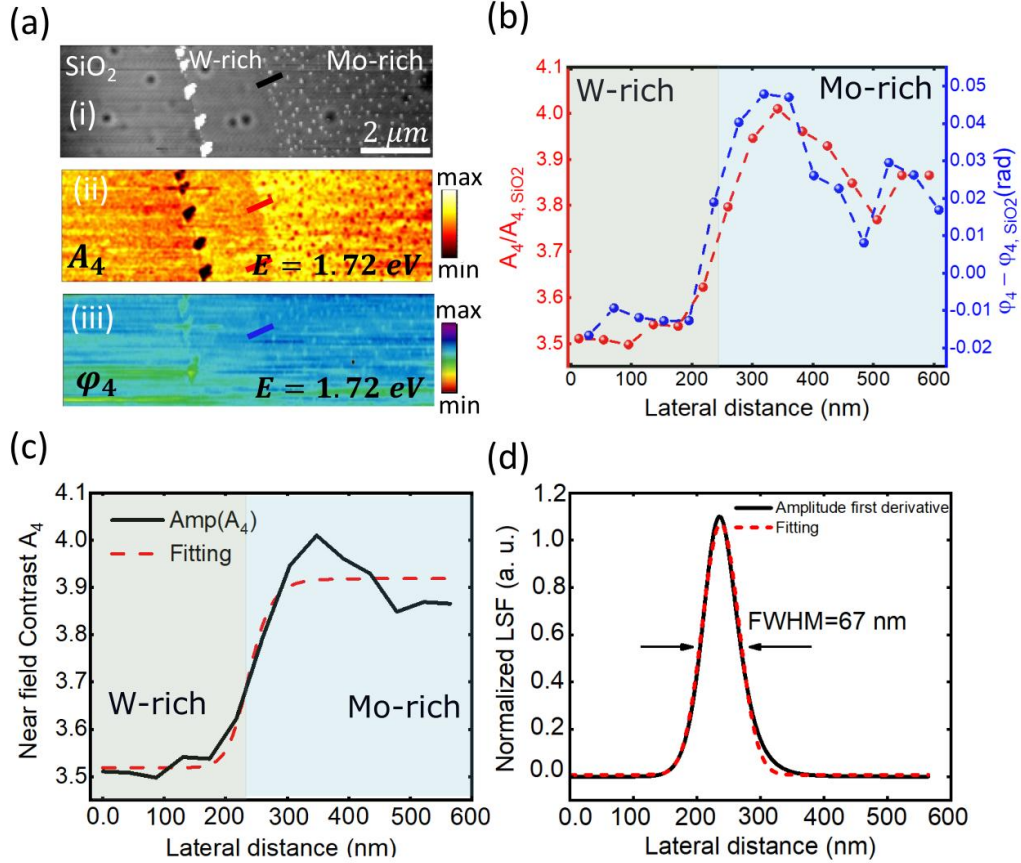

**Figure 4**. **High-resolution near-field imaging of the Mo–W lateral heterointerface.** (a) (iiii) AFM topography, fourth-harmonic near-field amplitude ($A_4$), and phase ($\varphi_4$) images acquired with enhanced spatial resolution across a selected region of the monolayer $Mo_xW_{1-x}S_2$ - $W_xMo_{1-x}S_2$ heterostructure at an excitation energy of 1.72 eV. These measurements are optimized to resolve the sharp interface between the W-rich and Mo-rich domains. (b) High-resolution line profiles of the fourth-harmonic near-field amplitude ($A_4$) and phase ($\varphi_4$), extracted across the heterointerface along the dashed line indicated in (a), showing an abrupt transition between the W-rich and Mo-rich regions. (c) Near-field amplitude $A_4$ line profile across the heterointerface (black solid line), together with a fit (red dashed line), used to quantitatively characterize the spatial variation of the near-field contrast at the interface. (d) First derivative of the near-field amplitude line profile (black symbols), fitted with a symmetric Lorentzian function (red dashed line) to determine the effective interface width and spatial resolution, yielding a full width at half maximum (FWHM) of approximately 67 nm.

# Tip-based PL spectroscopy of $Mo_xW_{1-x}S_2$ - $W_xMo_{1-x}S_2$ Heterostructure

To gain local insight into the optical properties sharp $Mo_xW_{1-x}S_2$ - $W_xMo_{1-x}S_2$ heterostructure and the interface region, we carried out tip-based point by point PL spectroscopy measurements at room temperature. Figure 5(a) displays a schematic of the experimental set-up, showing the laser excitation of 532 nm wavelength focused onto the AFM tip. The laser excitation power was fixed to 200 µW and the sample scanning step size (tip remains fixed) to 40 nm (which is about the diameter of the tip apex). Figure 5(b) shows the PL point spectra taken at three different regions of the heterostructure, from the W-rich area, from the interface and from inside the W-rich area. The PL spectra at the W-rich, interface, and Mo-rich regions were fitted using a sum of two Gaussian functions to account for the contributions of the neutral and bound exciton peaks in each region. For the Mo/W-rich side, the spectra were fitted with peaks corresponding to $X^0$($MoS_2$/$WS_2$) and $X^{BE}$($MoS_2$/$WS_2$). Two dominant peaks at the interface correspond to neutral excitons of $WS_2$ and $MoS_2$.. Figure 5(c) displays the hyperspectral AFM PL map of the band edge emission spatial profile of the $Mo_xW_{1-x}S_2$ heterostructure [17]. Figure 5(c)(i) is a 3D hyperspectral data cube taken by measuring an array of 60 by 60 pixels AFM PL spectra of the heterostructure monolayer. The x and y axes of the 3D data cube shown in Fig. 5(c) (i) indicate the plane of the sample surface while the z-axis corresponds to the photon energy axis (1.85 to 2.0 eV). The acquisition time for each spectrum was 1 sec, and the total acquisition time of 1 hour per image (see Materials and Methods for details). Panels (ii) and (iii) present equalized cut cross-section images extracted from the cube at two different energies. These maps visualize the spatial evolution of excitonic emission across the heterostructure. The corresponding non-equalized cut cross-section images are provided in Supplementary Figure SI3. These PL maps reveal distinct spatial variations in excitonic features across the heterostructure, highlighting how the neutral exciton intensities of both $MoS_2$-like and WS2-like regions evolve as the AFM tip traverses the interface. While Fig. 5(c) presents a 3D hyperspectral PL map acquired across the entire flake, Fig. 5(d) shows a 5 µm-long hyperspectral line scan taken perpendicular to the lateral interface, capturing the transition from W-rich to Mo-rich regions. This scan, with a 50 nm step size, reveals finer spatial evolution of PL spectra across the interface. The data are displayed as a 2D color map, with tip position along the x-axis and photon energy along the y-axis, illustrating energy-resolved excitonic features across the heterojunction. The inset shows normalized intensity is extracted from the 2D map in Fig. 5(d).

Panel e shows the imaginary part of the effective dielectric function of MoS$_2$/WS$_2$ alloys as a function of photon energy and Mo- filling fraction, calculated using Bruggeman effective medium theory. Distinct Lorentz oscillator resonances are visible. The inset displays the absorptance of a 6.2 A°-thick suspended film under p-polarized illumination at 45° incidence. The simulated absorptance map closely resembles the PL response in Fig. 5e, but here the variation is plotted against alloy filling factor instead of spatial position. We included this comparison to extend the experimental PL results in Fig. 5d–e with a theoretical model, showing that the observed excitonic trends across the heterostructure can be understood in terms of changes in the effective dielectric response with alloy composition. However, a well-resolved, spatially distinct interface (with clearly separated W-rich and Mo-rich regions), within 70 nm resolution, like that observed in the s-SNOM images in Fig. 3 and Fig. 4, was not possible. This is expected, mainly due to three reasons. (i) The strong plasmonic-induced dipolar response from the tip is for the most part oriented vertically, commensurate with charge accumulation, and thus produces electric-field (**E**) lines normal to the TMDC surface, very much like what is observed between the two plates of a capacitor. Indeed, as the tip becomes closer to the flat surface, it tends to short-circuit tip/TMDC interactions, meaning the tangential components of the electric field **E** tend to vanish very close to the tip. (The variation of this short-circuiting dependence with the tip's heigh, in contrast, is very accurately captured with tip demodulation methods. These are routinely utilized in near-field s-SNOM imaging for enhanced spatial resolution, but were otherwise not available in these PL measurements, so the total scattered PL field could not be filtered away from the tip.) On the other hand, from normal continuity of the displacement field **D**, the normal components of **E** on both sides of the air/TMDC interface are related by $E_{z,in} = E_{z,out}/\varepsilon_{TMDC}$. Given the large dielectric functions of both WS$_2$ and MoS$_2$, the normal electric field component $E_z$, responsible for inducing vertical polarization responses in the heterostructure, is not much amplified either. (ii) Here, the structure is illuminated by 532 nm light (2.33 eV) and the recollected luminescence intensity across the spectrum is shown. Considering the non-centrosymmetric character of few-layer WS$_2$ and MoS$_2$, the strength of this inelastic response can be described to scale with both some second-order dispersive nonlinear susceptibility $\chi^{(2)}(\mathcal{E}_{in} - \mathcal{E}_\Delta; \mathcal{E}_{in}, \mathcal{E}_\Delta)$ and the electric field magnitude at 532 nm. One can see in panel (e), however, that, the elastic response of the structure barely varies with the Mo-to-S ratio in the material at such wavelength, for which $\varepsilon_{WS_2} = 21.5 + 7.2i$ and $\varepsilon_{MoS_2} = 22.7 + 10.0i$ (the respective absorptances for WS$_2$ and MoS$_2$ are 0.036 and 0.049 in panel (e)).

Consequently, the PL response in this scenario will vary with frequency (energy) in as much as $\chi^{(2)}$ varies with $\varepsilon_\Delta$, $\varepsilon_{in}$ being fixed at 2.33 eV. (iii) Last, there is in this case a more mundane resolution-limiting factor: from panel (b), and also from Fig. 2(c), the widths of the two nonlinear resonances, peaked at roughly 1.87 and 1.99 eV (very close to the respective elastic excitons), are considerably large and, as a consequence, these two peaks are not resolved in energy (from the large imaginary part of $\varepsilon_{WS_2}$ and $\varepsilon_{MoS_2}$, long relaxation times—i.e. large losses—in at least one of the various scattering channels involved in $\chi^{(2)}$ are also expected). This lack of resolution is worsened by the fact that the Mo-rich region also presents, although quenched, another excitonic resonance at 2 eV, which somewhat overlaps with the W-rich exciton (see Fig3(e)).

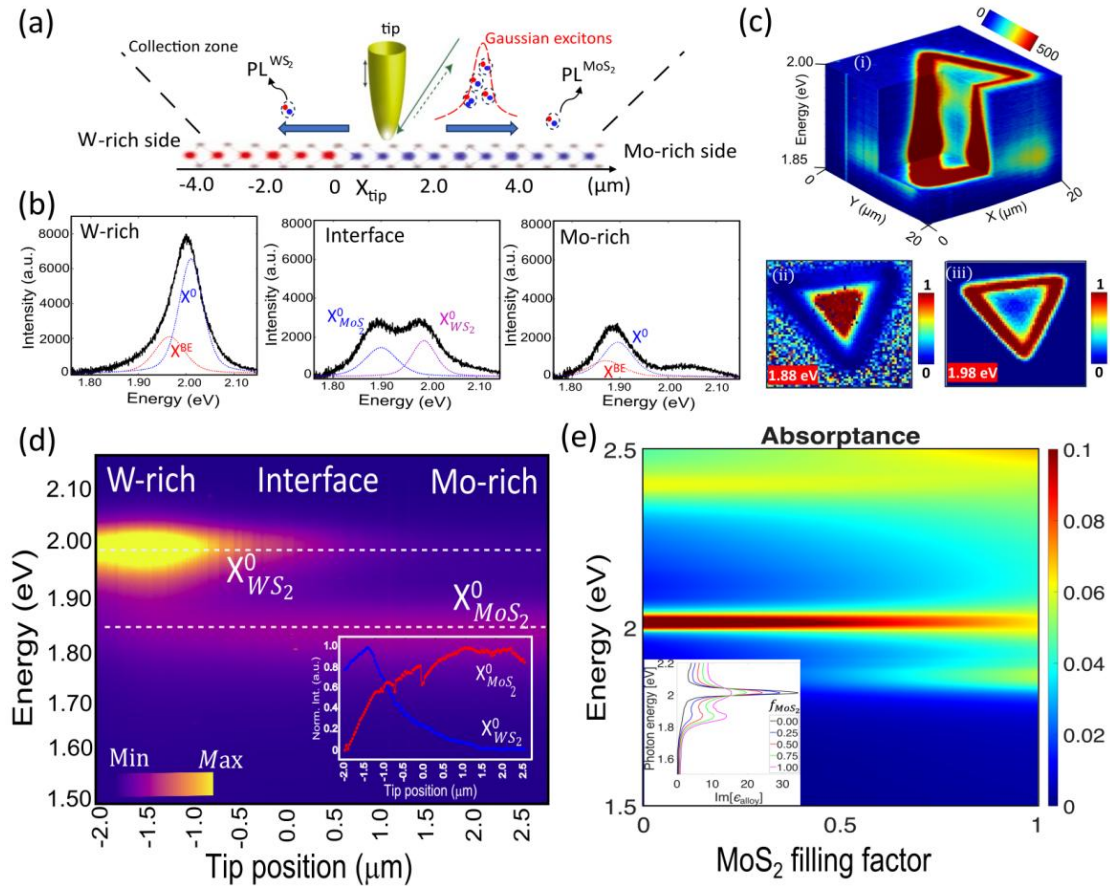

**Figure 5. AFM-PL mapping of the $Mo_xW_{1-x}S_2$ - $W_xMo_{1-x}S_2$ heterostructure.** (a) schematic of the lateral Heterostructure, AFM-PL measurement, and the resulting excitonic properties. (b) PL spectra taken across the Mo-rich side, interface, and W-rich side. Dashed lines are the fitted peak in each region. (c) 3D cube hyperspectral PL map of the $Mo_xW_{1-x}S_2$ - $W_xMo_{1-x}S_2$ heterostructure (i), the cut section images taken from the hyperspectral map at different energies (ii-iii). (d) PL energy map showing distinct spatial exciton energy variations across the heterostructure. The inset: Normalized extracted exciton energy peaks of d. (e) Absorptance map of a 6.2 A°-thick layer of TMDC suspended in air, under $p$ polarization and 45° oblique incidence, vs. energy and $MoS_2$

filling factor $f_{MoS_2}$ in the MoS$_2$/WS$_2$ alloy (Bruggeman effective-medium theory) at the interface. Resonances arise from Lorentz oscillators. The Inset: imaginary part of the effective dielectric function vs. energy, for different MoS$_2$ filling factors.

**Low-Temperature PL Spectroscopy at Different Excitation Powers**

Photoluminescence measurements at low temperature under varying excitation power provide insight into the exciton relaxation pathways driving the distinct band-edge features across the alloyed heterostructure. We performed power-dependent PL spectroscopy at 4 K on monolayer Mo$_x$W$_{1-x}$S$_2$ - W$_x$Mo$_{1-x}$S$_2$ in-plane heterostructures. Figure 6(a) shows spectra from the W-rich edge (region 3), Mo-rich center (region 1), and intermediate region (region 2), measured with excitation powers ranging from 10 μW to 2120 μW. PL spectra across all powers and regions were deconvoluted using 3–6 pseudo-Voigt peaks, shown as shaded components that best fit the experimental data.

To study the exciton relaxation pathways in different regions of the monolayer Mo$_x$W$_{1-x}$S$_2$ - W$_x$Mo$_{1-x}$S$_2$ in-plane heterostructure, we have fitted PL spectra of three different regions of the heterostructure. Figure 6(b) shows the peak positions for different fitted peaks in three different regions of the heterostructure monolayer as a function of excitation power to determine their origin. The deconvoluted PL spectrum of the Mo-rich center region exhibits three peaks, with the highest-energy feature (~2.05 eV) remaining weak across all excitation powers. Peak positions in this region display negligible shifts within the uncertainty range of spectral fitting. In contrast, the W-rich edge region also shows three peaks, with excitation power inducing shifts of ~5 meV for peaks 1 and 2, and ~10 meV for peak 3. The intermediate MoW region reveals six peaks, consistent with a superposition of spectral features associated with both MoS$_2$- and WS$_2$-rich domains. Across all regions, the fitted peak positions exhibit minimal dependence on excitation power. The modest blue shifts observed are attributed primarily to quasi-Fermi level splitting under increased photoexcited carrier densities[34, 35]. Competing many-body effects, such as strain-induced bandgap renormalization and exciton binding energy reduction, may drive redshifts[36, 37] but are not dominant in our measurements.

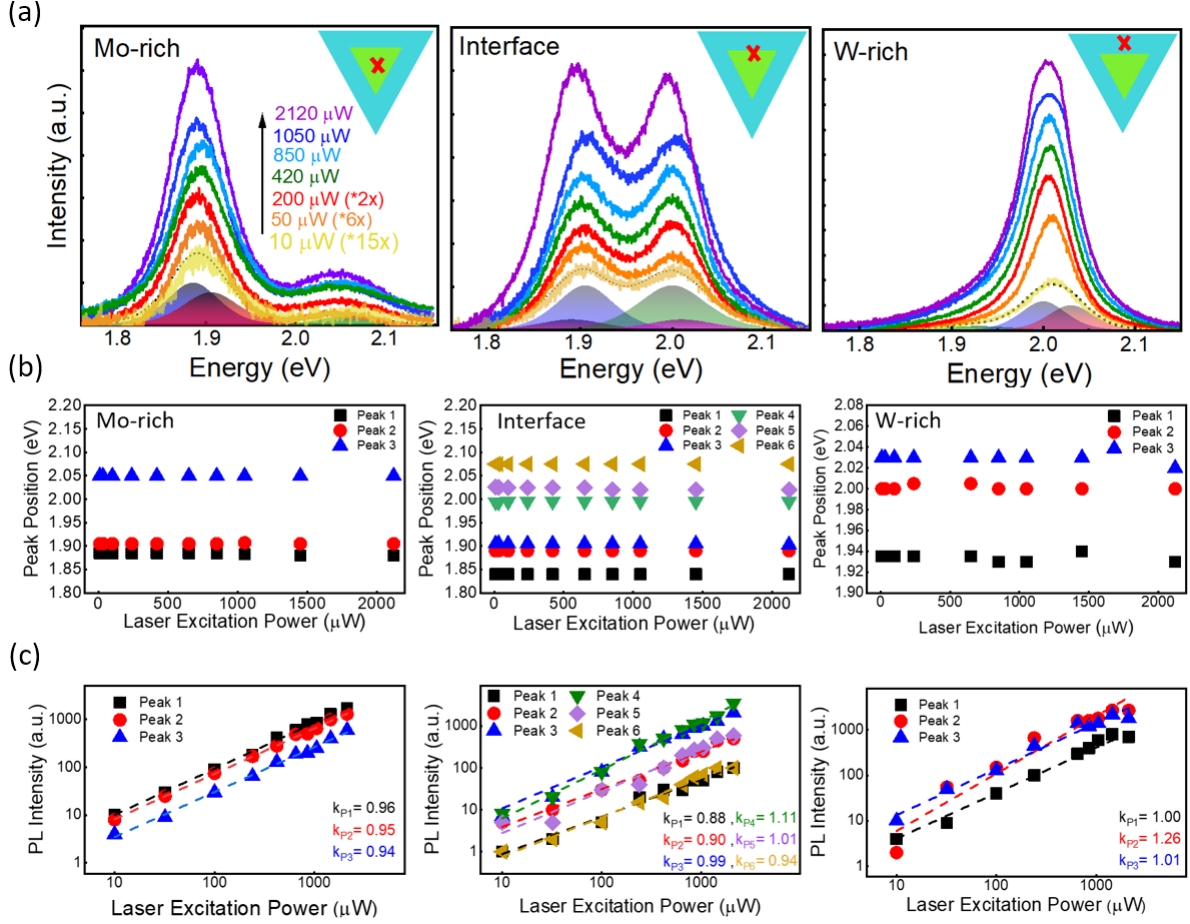

**Figure 6. Excitation laser power dependence of the PL spectra for the $Mo_xW_{1-x}S_2$ - $W_xMo_{1-x}S_2$ heterostructure taken at T = 4 K.** (a) PL spectra acquired from three regions: the center (Mo-rich), the interface (Mo-W mixed region), and the edge (W-rich). (b) Power dependence of the PL peak positions at these three regions. (c) PL intensity plots of peaks 1–3 as a function of excitation power for each region.

To elucidate the origin of the observed PL peaks across different regions of the heterostructure, we analyzed the excitation power dependence of their integrated intensities using the power law $I=P^k$, where k characterizes the recombination mechanism. As shown in Fig. 6(c), the extracted k-values indicate nearly linear behavior (k≈1) for all peaks, suggesting excitonic origin. In the Mo-rich region, peaks at ~2.05 eV and ~1.905 eV are assigned to B and A excitons, respectively, while the ~1.88 eV peak corresponds to a bound exciton with a binding energy of ~22 ± 2 meV, consistent with prior reports[34, 35]. In the W-rich region, the ~2.02 eV peak is attributed to the A exciton, and the ~1.26 eV peak to a bound exciton with a binding energy of ~20 ± 4 meV, again in agreement with literature [36, 37]. A lower-energy peak (~1.00 k-value) is identified as a defect-bound exciton. In the interface region, the presence of six peaks reflects a combination of transitions associated with both Mo- and W-rich domains. Their near-linear power dependence confirms excitonic recombination, including contributions from bound and defect-mediated excitons, such as P1 (~1.84 eV) and P6 (~2.07 eV), which are attributed to impurity-bound transitions.

**Evolution of PL Spectra as a Function of Temperature**

To gain deeper insight into excitonic behavior of the $Mo_xW_{1-x}S_2$ - $W_xMo_{1-x}S_2$ in-plane heterostructures, we performed temperature dependent PL measurements across different regions of the sample. Fig. 7(a) presents the evolution of PL spectra for three representative areas of the in-plane heterostructure, $Mo_xW_{1-x}S_2$ - $W_xMo_{1-x}S_2$, over the temperature range of 4 K to 300 K. All spectra were acquired using a 532 nm excitation laser at a fixed power of 200 µW, with an exposure time of 2 seconds. As the temperature increases, a redshift in the optical transition energies (i.e., PL peak positions) is observed, consistent with the typical temperature-induced bandgap narrowing. Fig. 7(b) illustrates the temperature dependence of the PL peaks across the three regions. To quantitatively describe the shift of peaks near the band edge (peaks 1–3), we employed the O'Donnell model [38] to fit the experimental data. The resulting fits, indicated by the dot-dash lines in Fig. 7(b) for three different regions of the heterostructure.

$$E_g(T) = E_g(0) - S<\hbar\omega>\left[coth\left(\frac{<\hbar\omega>}{2kT}\right) - 1\right] \qquad (1)$$

where $E_g(0)$ represents the ground-state transition energy at 0 K, S is a dimensionless exciton-phonon coupling constant, and $<\hbar\omega>$ denotes the average phonon energy. The fitting parameters extracted for each region of the heterostructure are summarized in Table SI1. According to the fitting parameters the average phonon energy $<\hbar\omega>$ associated with the A-exciton in the Mo-rich side is found to be 50 meV in good agreement with previously reported values[34]. Similarly, for the A-exciton on the W-rich side, the extracted phonon energy ranges from ~35 to 44 meV, consistent with earlier studies[37, 39]. According to Fig. 7(b) the B-exciton peak on the Mo-rich side exhibits a redshift of approximately 50 meV as the temperature increases from 4 K to 300 K, aligning well with literature values. For the A-exciton across all three regions, the observed redshift over the same temperature range is around 33±3 meV.

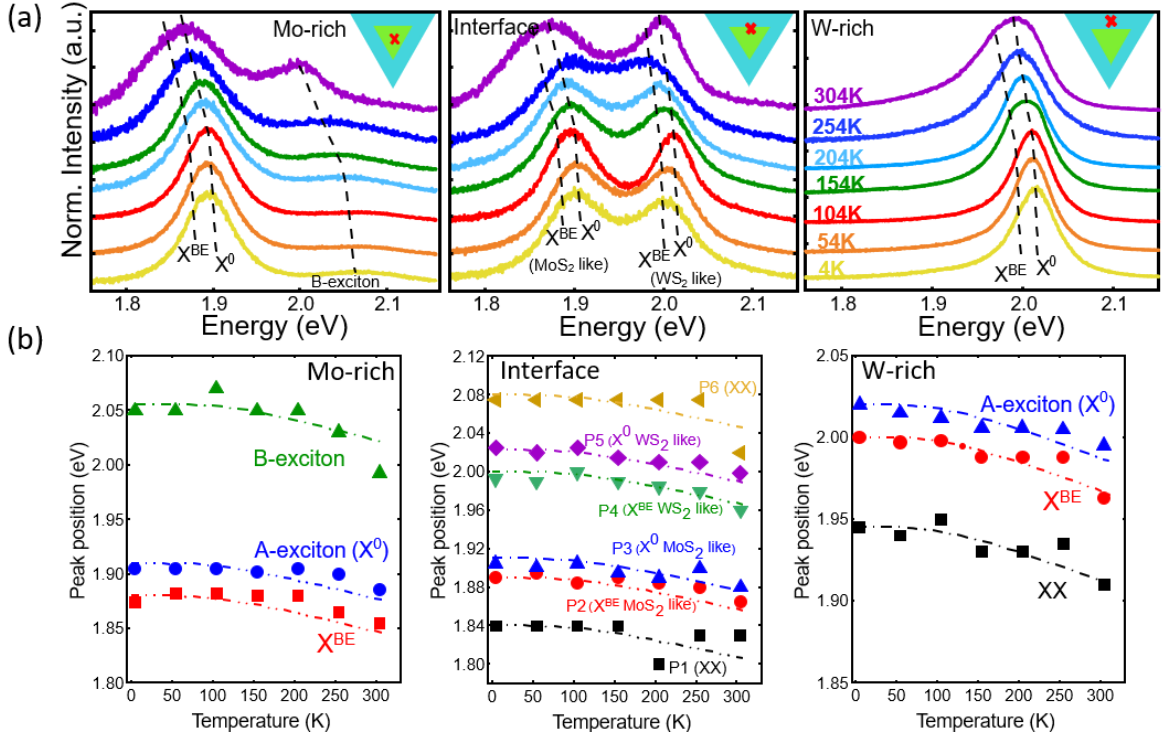

**Figure 7. Temperature dependent evolution of PL Spectra in $Mo_xW_{1-x}S_2$ monolayer heterostructure.** (a) shows PL Spectra as a function of temperature in different regions; region1 (Mo-rich side), region 2 (intermediate MoW side), and region3 (W-rich side) of the sample. (b) Temperature dependence of the fitted PL peak positions for the three regions of the $Mo_xW_{1-x}S_2$ - $W_xMo_{1-x}S_2$ in-plane heterostructures. The dash-dot lines represent fits based on the conventional temperature dependence of semiconductor bandgaps.

**CONCLUSION**

In summary, using a liquid-phase precursor-assisted synthesis approach, we synthesized $Mo_xW_{1-x}S_2$ - $W_xMo_{1-x}S_2$ monolayer in-plane heterostructures featuring a sharp nanoscale lateral interface. The structural and optical properties of the heterojunction were comprehensively characterized using AC-HRSTEM and scattering-type near-field optical microscopy (s-SNOM). AC-HRSTEM enabled observation of distinct compositional domains revealing the center and edge regions are predominantly $MoS_2$ and $WS_2$ with a small amount of W and Mo dopants, respectively, and the interface region is alloyed $Mo_xW_{1-x}S_2$. s-SNOM imaging provided nanoscale amplitude and phase contrast maps, resulting from differences in the local dielectric function between Mo-rich and W-rich regions across the nanointerface. AFM-PL spectroscopy was performed to study diffraction limited spatial variations in PL peak positions and intensities of neutral, and bound excitons across the heterostructure. Power- and temperature-dependent PL spectra further highlighted the evolution excitonic features across different regions, and demonstrated reduced sulfur monovacancy density compared to previously studied alloyed structures. These results demonstrate that in-plane TMDC heterostructures with atomically defined interfaces offer a robust platform for nanoscale optical control and excitonic engineering, paving the way for future applications in next-generation optoelectronic and quantum optical devices.

## MATERIALS AND METHODS:

**The Growth of $Mo_xW_{1-x}S_2$ - $W_xMo_{1-x}S_2$ In-Plane Heterostructures.** Monolayer $Mo_xW_{1-x}S_2$ - $W_xMo_{1-x}S_2$ in-plane heterostructures were synthesized by a liquid-phase precursor-assisted approach using ammonium metatungstate hydrate (AMT) and ammonium heptamolybdate (AHM) as W- and Mo-precursors, respectively, and using sodium cholate hydrate (SC) as a growth promoter. The mixed aqueous solution containing AMT, AHM, and SC was first spin-coated onto a $SiO_2$/Si substrate. Then a high-temperature sulfurization process (ramped to 700 °C and held for 10 min, and then further ramped to 800 °C and held for another 10 min) was carried out, using an argon flow of 100 sccm as a carrier gas.

**AC-HAADF-STEM.** As-grown $Mo_xW_{1-x}S_2$ - $W_xMo_{1-x}S_2$ in-plane heterostructures were transferred onto Quantifoil gold grid using conventional PMMA-based wet chemical transfer . [40] For AC-HAADF-STEM imaging, a double spherical aberration corrected FEI Titan[3] G2 S/TEM 60-300 operated at 80 kV using a high angle annular dark field (HAADF) detector with a collection angle of 42-244 mrad, camera height of 115 mm, and convergence angle of 30 mrad. A low voltage and beam current (46 pA) was used at high resolution to decrease irradiation damage. To enhance visibility and reduce noise in STEM images, all acquired high-resolution images were processed by Gaussian Blur filter (radius = 2.00) using ImageJ software.

**Optical Measurements.** The power and temperature-dependent PL spectra were measured by the confocal laser scanning microscope system equipped with a vibration-free closed-cycle cryostat (Attodry 800, attocube). A 532 nm CW laser as an excitation source was focused into a small spot with a diameter of approximately 2-3 μm on the sample through a 100× objective lens (APO/VIS, N.A. = 0.82; attocube) inside the vacuum chamber. The PL spectra was then collected by the same lens and filtered the excitation signal by a 532 nm long-pass filter before entering a spectrometer (Andor) which consisted of a monochromator and a thermoelectrically cooled CCD camera. For SNOM Imaging of $Mo_xW_{1-x}S_2$ - $W_xMo_{1-x}S_2$ in-plane heterostructures were performed using a commercial Visible scattering-type scanning near-field microscope (s-SNOM) from Attocube co. s-SNOM was based on a tapping mode AFM equipped with a cantilevered metal-coated tip. The tip oscillates at a resonance frequency $\Omega \approx 270$ kHz and a tapping amplitude of $\approx 100$ nm. A 532 nm CW was focused by a parabolic mirror onto the metalized tip and interacts with the sample. The scattered light from the tip-sample junction was demodulated at the fourth harmonics of the tip resonance frequency and detected by phase modulation (pseudo-heterodyne) interferometry, resulting in simultaneous topography, optical amplitude, and phase images.

## AUTHOR INFORMATION

**Corresponding Author**

Yohannes Abate (*yohannes.abate@uga.edu*)

**Author Contribution**

Y.A. conceived the idea and guided the overall project. M.G. carried out the s-SNOM and AFM-PL experiments and analyzed the experimental data. T.Z., and D.Z grew the sample and AC-HRSTEM measurements. M.G., S.G, M.A, S.S helped with measurements and analysis of data. A.G.R and D.M.S helped with the PL simulations. All authors discussed the results and contributed to writing the manuscript.

**Notes**

The authors declare no competing financial interest.

## ACKNOWLEDGMENT

The work of M.G. and S.S. is supported by the Gordon and Betty Moore Foundation, grant DOI 10.37807/GBMF12246 and the work of M. A. is supported by the Air Force Office of Scientific Research (AFOSR) grant number FA9550-23-1-0375. Partial support for S.G. comes from the National Science Foundation (NSF) Grant No. 2152159 (NRT-QuaNTRASE). D.M.S acknowledges support from a Ramón y Cajal fellowship (Grant No. RYC2023-045265-I) funded by MICIU/AEI/10.13039/501100011033 and by the ESF+, as well as from Xunta de Galicia Regional Government (Consolidation of Competitive Research Units type C: Grant No. ED431F 2025/22).

# Supplimentary data

TableS 1. Fitted Values of the Exciton−Phonon Coupling Strength, S, the Average Phonon Energy, $\langle \hbar\omega \rangle$, and $E_{g(0)}$ of neutral A exciton, $Trion, B\ exciton, Bound\ exciton$ for different regions of heterostructure $Mo_xW_{1-x}S_2$ monolayer.

|  | Mo-rich side of heterostructure $Mo_xW_{1-x}S_2$ | MoW side of heterostructure $Mo_xW_{1-x}S_2$ | W-rich side of heterostructure $Mo_xW_{1-x}S_2$ |
|---|---|---|---|
| $E_{0B}$ (eV) | 2.055 | - | - |
| $E_{0A}$ (eV) | 1.91 | 1.91-2.025 | 2.025 |
| $E_{0T}$ (eV) | 1.88 | 1.89-2.00 | 1.995 |
| $E_{0XX}$ (eV) | 1.862 | 1.84-2.08 | 1.945 |
| $\langle\hbar\omega\rangle$ (meV) | 30 | 31 | 30 |
| S | 1.2 | 1.2 | 1.3 |

**Figure SI1.** 2D image of Temperature dependence PL spectra of the $Mo_xW_{1-x}S_2$-$W_xMo_{1-x}S_2$ heterostructure.

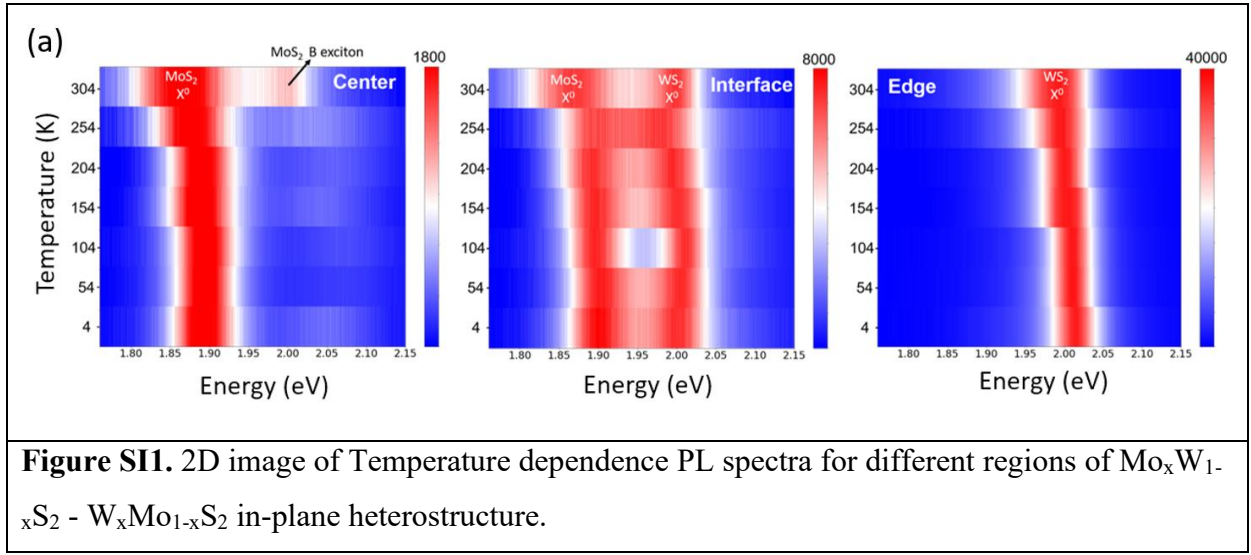

**Figure SI1.** 2D image of Temperature dependence PL spectra for different regions of $Mo_xW_{1-x}S_2$ - $W_xMo_{1-x}S_2$ in-plane heterostructure.

**Temperature dependence of FWHM for $Mo_xW_{1-x}S_2$ - $W_xMo_{1-x}S_2$ in-plane heterostructures.**

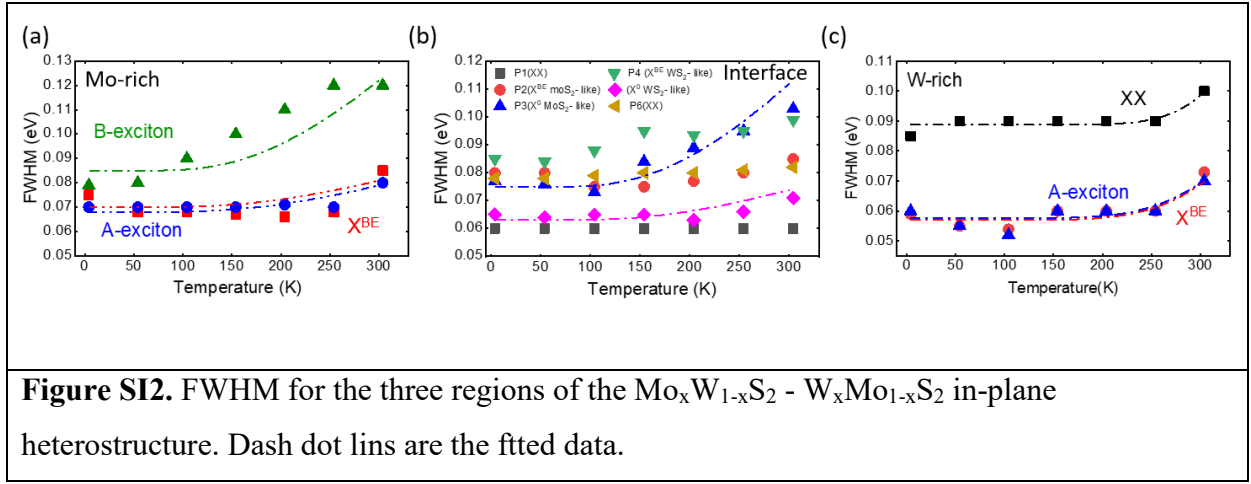

**Figure SI2.** FWHM for the three regions of the $Mo_xW_{1-x}S_2$ - $W_xMo_{1-x}S_2$ in-plane heterostructure. Dash dot lins are the ftted data.

The temperature dependence of the exciton linewidth in TMDCs is commonly attributed to electron–phonon interactions, and can be described by the following equation:

$$\gamma = \gamma_I + \frac{b}{exp\left(\frac{\Theta}{KT}\right) - 1} \quad (1)$$

Where $\gamma_I$ represents the temperature-independent inhomogeneous broadening, and $\Theta$ denotes either the energy of a dominant phonon mode or an average phonon energy. We used the equation (1) to fit the FWHM of experimental data for different peaks at three different regions of the ahetetostructure as shown the dashed dots in figure SI2.

**2D PL map of the $Mo_xW_{1-x}S_2$ - $W_xMo_{1-x}S_2$ in-plane heterostructure.**

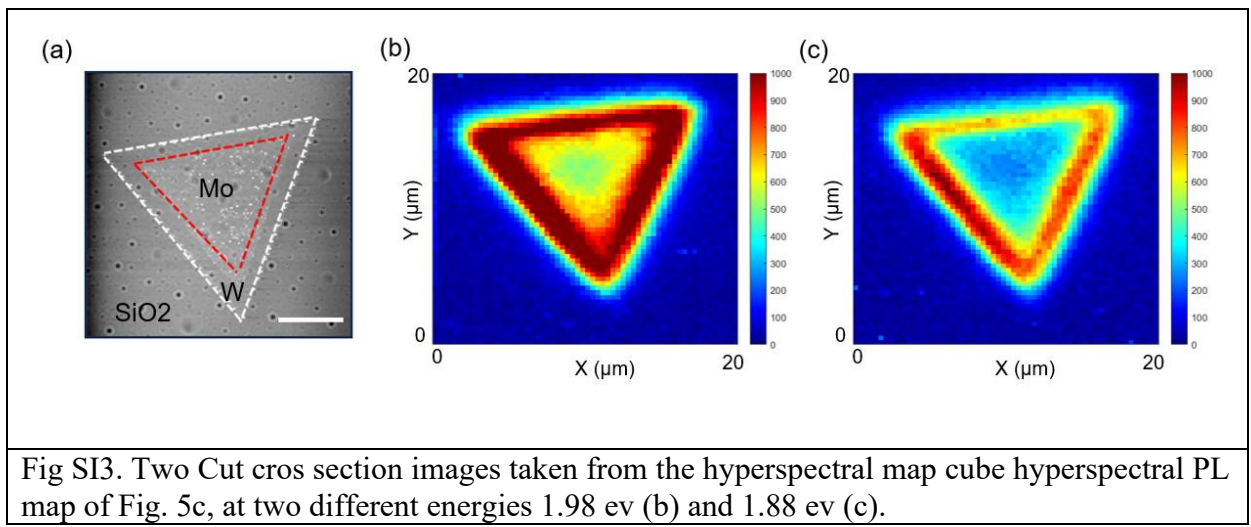

Fig SI3. Two Cut cros section images taken from the hyperspectral map cube hyperspectral PL map of Fig. 5c, at two different energies 1.98 ev (b) and 1.88 ev (c).